\documentstyle[11pt]{article}                                     
\newcommand{\frak}[1]{{\mathbf #1}}                                   

\def\AFOUR{%
\setlength{\textheight}{9.0in}%
\setlength{\textwidth}{5.75in}%
\setlength{\topmargin}{-0.375in}%
\hoffset=-.5in%
\renewcommand{\baselinestretch}{1.17}%
\setlength{\parskip}{6pt plus 2pt}%
}
\AFOUR                                           
\def\car{\mathop{\square}}
\def\carre#1#2{\raise 2pt\hbox{$\scriptstyle #1$}\car_{#2}}

\makeatletter
\def\section{\@startsection {section}{1}{\z@}{-3.5ex plus -1ex minus
 -.2ex}{2.3ex plus .2ex}{\large\bf}}
\def\subsection{\@startsection{subsection}{2}{\z@}{-3.25ex plus -1ex minus
 -.2ex}{1.5ex plus .2ex}{\normalsize\bf}}
\makeatother
\newcommand{\nc}{\newcommand}
\newcommand{\rnc}{\renewcommand}
\nc{\be}{\begin{equation}}
\nc{\ee}{\end{equation}}
\nc{\bea}{\begin{eqnarray}}
\nc{\eea}{\end{eqnarray}}

\def\slash#1{\setbox0=\hbox{$#1$}#1\hskip-\wd0\hbox to\wd0{\hss\sl/\/\hss}}

\def\href#1#2{{#2}}

\rnc{\a}{\alpha}
\nc{\ab}{\bar{\a}}
\nc{\ap}{\a^{+}}
\nc{\abm}{\ab^{-}}
\rnc{\b}{\beta}
\nc{\bb}{\bar{\b}}
\nc{\bbp}{\bb_{\zb}^{+}}
\nc{\bm}{\b_{z}^{-}}
\nc{\oa}{\overline{\a}}
\nc{\ob}{\overline{\b}}
\rnc{\gg}{\gamma}
\rnc{\d}{\delta}
\nc{\f}{\phi}
\nc{\fb}{\bar{\phi}}
\nc{\vf}{\varphi}
\nc{\p}{\psi}

\rnc{\c}{\chi}
\nc{\la}{\lambda}
\nc{\m}{{\mathrm m}}
\nc{\n}{\nu}
\rnc{\o}{\omega}
\nc{\Om}{\Omega}
\rnc{\t}{\theta}
\nc{\eps}{\epsilon}
\rnc{\S}{\Sigma}
\nc{\F}{\Phi}
\nc{\trac}[2]{{\textstyle\frac{#1}{#2}}}
\nc{\ex}[1]{\mbox{e}^{\,\textstyle#1}}
\nc{\mat}[4]{\left(\begin{array}{cc}#1&#2\\#3&#4\end{array}\right)}
\nc{\som}[9]{\left(\begin{array}{ccc}#1&#2&#3\\#4&#5&#6\\#7&#8&#9%
\end{array}\right)}
\nc{\tr}{\mathop{\mbox{tr}}\nolimits}
\nc{\ad}{\mathop{\mbox{ad}}\nolimits}
\nc{\Tr}{\mathop{\mbox{Tr}}\nolimits}
\nc{\Det}{\mathop{\mbox{Det}}\nolimits}
\nc{\rk}{\mathop{\mbox{rk}}\nolimits}
\nc{\ra}{\rightarrow}
\nc{\Ra}{\Rightarrow}
\nc{\LRa}{\Leftrightarrow}
\nc{\ot}{\otimes}
\rnc{\ss}{\subset}
\nc{\nul}{\noindent\underline}
\nc{\non}{\nonumber\\}
\nc{\subs}[1]{{\vspace*{0.5cm}}%
{\noindent\underline{#1}}{\addcontentsline{toc}{subsection}{#1}}%
{\vspace*{0.3cm}}}
\nc{\zb}{\bar{z}}
\rnc{\lg}{\frak{g}}
\nc{\lt}{\frak{t}}
\nc{\lk}{\frak{k}}
\nc{\lh}{\frak{h}}
\nc{\pik}{\Pi_{\lk}}
\nc{\pip}{\Pi_{+}}
\nc{\pim}{\Pi_{-}}
\nc{\pih}{\Pi_{\lh}}
\nc{\jz}{J_{z}}
\nc{\jzh}{\jz^{\lh}}
\nc{\jzp}{\jz^{+}}
\nc{\jzm}{\jz^{-}}
\nc{\del}{\partial}
\nc{\dz}{\del_{z}}
\nc{\dzb}{\del_{\bar{z}}}
\nc{\az}{A_{z}}
\nc{\azb}{A_{\bar{z}}}
\nc{\g}{g^{-1}}
\nc{\dw}{\Delta_{W}}
\nc{\Ad}{{\mbox{Ad}}}
\nc{\ks}{Ka\-za\-ma-\-Su\-zu\-ki}
\nc{\KS}{\ks}
\nc{\ksm}{\ks\ model}
\rnc{\AA}{{\Bbb A}}
\nc{\BB}{{\Bbb B}}
\nc{\CC}{{\Bbb C}}
\nc{\PP}{{\Bbb P}}
\nc{\cpm}{\CC\PP(m)}
\nc{\cpn}{\CC\PP(n)}
\nc{\cp}[1]{\CC\PP(#1)}
\nc{\gmn}{G(m,m+n)}
\nc{\gmnk}{\gmn_{k}}
\nc{\cO}{{\cal O}}
\nc{\bcO}{\bar{\cO}}
\nc{\bO}{\bar{O}}
\nc{\oQ}{\overline{Q}}
\nc{\ie}{{\it i.e.~}}
\nc{\eg}{{\it e.g.~}}
\begin{document}
\global\parskip=4pt
\makeatother\begin{titlepage}
\begin{flushright}
{hep-th/98mmnnn}
\end{flushright}
\vspace*{0.5in}
\begin{center}
{\LARGE\sc  On the perturbative corrections
around D-string instantons }\\
\vskip .3in
\makeatletter

\begin{tabular}{c}
{\bf E. Gava}, \footnotemark 
\\ 
INFN, ICTP and SISSA, Trieste, Italy, \\
\end{tabular}

\begin{tabular}{c}
{\bf A. Hammou}, \footnotemark 
\\ 
SISSA, Trieste, Italy, \\
\end{tabular}

\begin{tabular}{c}
{\bf J.F. Morales}, \footnotemark
\\ 
INFN, Sezione di ``Tor Vergata'', Roma, Italy,  \\
\end{tabular}

\begin{tabular}{c}
{\bf K.S. Narain}\footnotemark
\\ ICTP, P.O. Box 586, 34014 Trieste, Italy\\
\end{tabular}

\end{center}
\addtocounter{footnote}{-3}%
\footnotetext{e-mail: gava@he.sissa.it}
\addtocounter{footnote}{1}%
\footnotetext{e-mail: amine@sissa.it}
\addtocounter{footnote}{1}%
\footnotetext{e-mail: morales@sissa.it}
\addtocounter{footnote}{1}%
\footnotetext{e-mail: narain@ictp.trieste.it}
\vskip .50in
\begin{abstract}
\noindent 
We study ${\cal F}^4$-threshold corrections in an eight dimensional
S-dual pair of string theories, as a prototype of dual string vacua with sixteen
supercharges. We show that the orbifold CFT description of
D-string
instantons gives rise to a perturbative expansion similar to the one
appearing on the fundamental string side. 
By an explicit calculation, using the Nambu-Goto action in the static gauge, we show
that the first subleading term agrees precisely on the two sides. We then give a
general argument to show that the agreement extends to all orders.

\end{abstract}
\makeatother
\end{titlepage}
\begin{small}
\end{small}

\setcounter{footnote}{0}

Under the type I/heterotic duality map, length scales are 
related by $L_h=L_I/\sqrt{\lambda_I}$, with $\lambda_I$ the type I
string coupling constant and $L_h$, $L_I$ heterotic and type I
scales, respectively. 
This implies, in particular, that an expansion in the large volume 
limit of a toroidal compactified heterotic theory can be translated to
a genus expansion on the type I side.
In \cite{bfk} the exact moduli depence of ${\cal F}^4$ and ${\cal R}^4$
terms in the low energy effective actions for toroidal compactifications
of heterotic string to $D\geq 5$ dimensions was computed. 
These terms are special in the sense that they are related 
by supersymmetry to 
CP odd couplings, for which only one-loop perturbative corrections 
are possible \cite{yasuda}. It is therefore expected that
this is also the case for their supersymmetric partners. 
Moreover, for $D\geq 5$, the only non-perturbative object
on the heterotic side, which is the fivebrane instanton, 
is always infinitely 
massive to give contribution. Therefore one expects that 
the one-loop result is exact even non-perturbatively.
The one-loop formula can be expressed, in the type I variables, 
as a sum of a finite number of perturbative corrections 
and an infinite series of D-string instanton contributions.
Indeed, for $D\geq 5$, the D-string instanton is the only
source of non-perturbative contributions on the type I side.
The form of the N-instanton contribution in this sum suggests 
that it can be read off from the $O(N)$ gauge theory describing N
nearby D-strings, whose worldsheets are wrapped on a two-torus
\cite{bv}. More precisely, the leading behaviour
in the volume of the compactification torus for 
the $SO(32)$ ${\cal F}^4$
and ${\cal R}^4$ gauge and gravitational couplings respectively
were shown in 
agreement \cite{bv} with the exact formula found in a
perturbative computation in the heterotic side \cite{bfkov}. Similar
results were found in \cite{bgmn}, where the 
moduli dependence for $O(2,2)$ ${\cal F}^4$ gauge
couplings, for the gauge fields coming from the KK 
reduction of the metric and antisymmetric tensor,
were considered in a class of orientifold models with 16 supercharges. 
Also, higher derivative
couplings involving graviphotons have been recently discussed in
\cite{fs}. In all the cases, the
D-string instanton coupling to
the gauge fields are obtained from the classical 
D-string instanton action, while
the contributions from quantum fluctuations around the
D-string instanton background are encoded in the elliptic genus of the
corresponding $O(N)$ gauge theory. 

The perturbative 
computations of \cite{bfkov, bgmn} show
that, on top of the leading D-string instanton corrections, there are 
a finite number of contributions corresponding to perturbative
corrections in the D-string instanton background. 
One can ask the question whether
the CFT describing the infrared limit of the 
D-string gauge theory captures
some information about these subleading terms. 
The aim of this letter is to show that this is indeed the case.
We will perform our analysis in the simplest context of the
eigth dimensional dual pair obtained via an orbifold/orientifold
of type IIB theory considered in \cite{bgmn}. 
The dual orbifold/orientifold actions are defined
by the orbifold $(-)^{F_L}\sigma_V$ (``fundamental side'')
and orientifold $\Omega\sigma_V$ (``type I side'')    
$Z_2$ actions, with $\sigma_V$ a shift of order two in the two
torus. 

Let us first start by briefly reviewing the computation of the moduli
dependence of ${\cal F}^4$ couplings in the the fundamental side
and how the result agrees, at the leading order in a large volume expansion,
with what one gets from the CFT describing the D-string system.
We will subsequently show that the CFT description also gives rise to a
perturbative expansion which has the same structure as the one appearing on the 
fundamental string side. In particular, we will show a perfect
agreement for the first subleading correction.

The  ${\cal F}^4$ couplings we are intersted in are obtained from
the one-loop string amplitudes: 
\bea
{\cal A}_\ell=\langle (V^L_8)^\ell(V^L_9)^{4-\ell} \rangle
\label{Aa}
\eea    
with
\be
V^L_i = \int d^2 z (G_{\mu i}+i B_{\mu i})(\partial X^{\mu}-\frac{1}{4} 
p_{\nu} S\gamma^{\mu \nu}S)(\bar{\partial} X^{i}-\frac{1}{4} 
p_{\rho} \tilde{S}\gamma^{i \rho}\tilde{S})\label{gaugevertex}  
e^{i p X}  
\ee 
the vertex operators of the left $O(2,2)$ gauge fields arising
from the $F_i\equiv G_{\mu i}+iB_{\mu i}$ 
components of the metric and antisymmetric tensor.
Here and in the following we will denote with capital indices
$M=(\mu,i)$ the ten dimensional noncompact $\mu=0,1,\dots 7$ 
and compact $i=8,9$ directions.    
The combinations
with the minus sign represent the graviphoton vertex operators,
which carry always additional power of momenta and are therefore
irrelevant for the computation of ${\cal F}^4$ couplings. 
It is convenient to define a generating function $Z(\nu_i,\tau,\bar\tau)$,
in term of which the above amplitudes read:
\be
{\cal A}_\ell = 
t_8 F_8^\ell F_9^{4-\ell}\int_{\cal F} \frac{d^2 \tau}{\tau_2^2}
\tau_2^4 \frac{\partial^\ell}{\partial\nu_8^\ell}
\frac{\partial^{4-\ell}}{\partial\nu_9^{4-\ell}}Z(\nu_i,\tau,\bar\tau)
\label{ampl}
\ee
with ${\cal F}$ the fundamental domain for 
the modulus of world-sheet torus.  
$Z(\nu_i,\tau,\bar\tau)$ is the partition function arising from
a perturbed Polyakov action, whose bosonic part is: 
\be
S({\nu_i}) = \frac{2\pi}{\alpha^\prime} \int d^2 \sigma ( \sqrt{g}
  G_{\mu\nu}g^{\alpha\beta}\partial_\alpha X^\mu \partial_\beta X^\nu
+i  B^{^{NS}}_{\mu\nu}\varepsilon^{\alpha\beta}
\partial_\alpha X^\mu \partial_\beta X^\nu+
\sqrt{g}\frac{\alpha^\prime}{2\pi\tau_2}\nu_i\bar{\partial}X^i ),
\label{polyakov}
\ee  
and
$\bar{\partial}=
\frac{1}{\tau_2}(\partial_{\sigma_2}-\tau\partial_{\sigma_1})$.
The geometry of the target torus are described as usual by the two
complex moduli
\bea
 T &=& T_1 + iT_2 = \frac{1}{\alpha' } (B^N_{89} + i\sqrt{G})\nonumber\\
 U &=& U_1 + i U_2 = ( G_{89} + i\sqrt{G} )/ G_{88} \ ,
\eea
where $G_{ij}$ and $B^N_{ij}$  are the
$\sigma$-model metric and NS-NS antisymmetric
tensor. Since the theory we are considering involves half shift along $X^8$
direction, it is convenient to replace  the radius $R^8$ by $2 R^8$.
Indeed this new radius is what appears on the dual $IIB/\Omega \sigma_V$ theory
(upto the scaling by the string coupling constant). As a result
$T$, $U$ and $\nu_8$ are replaced by $2T$, $U/2$ and $2\nu_8$ respectively and
furthermore all
the windings along $\sigma_1$ and $\sigma_2$ are integer valued.

 The partition function $Z(\nu_i,\tau,\bar\tau)$ is given by 
\be
\int_{\cal F}\frac{d^2 \tau}{\tau_2^2} Z(\nu_i,\tau\bar\tau) =
\frac{{\cal V}_8}{(4\pi^2 \alpha^\prime)^4}\int_{\cal F}
\frac{d^2 \tau}{\tau_2^2} 
\sum_{\epsilon}\Gamma^{\epsilon}_{2,2}(\nu_i){\cal A}_\epsilon
\label{Zl}
\ee
with 
\be
\Gamma^{\epsilon}_{2,2}(\nu_i)=
\frac{4T_2}{U_2}\,
 \sum_{W,{\epsilon}}
 e^{ 2\pi i  T {\rm det}W }
e^{- \frac{\pi T_2 }{ \tau_2 U_2 }
\big| (1\; U)W  \big( {\tau \atop -1} \big) \big| ^2}
e^{- \frac{\pi}{ \tau_2}
(\nu_8\,\nu_9)W \big( {\tau \atop -1} \big) }
\label{lattice}
\ee
and ${\cal A}_\epsilon$ the anti-holomorphic BPS partition functions 
\bea
{\cal A}_{+-}\equiv
\frac{1}{2}\frac{\vartheta_2(\bar{q})^4}{\eta(\bar{q})^{12}}
\quad , \quad
{\cal A}_{-+}\equiv
\frac{1}{2}\frac{\vartheta_4(\bar{q})^4}{\eta(\bar{q})^{12}}
\quad , \quad
{\cal A}_{--}\equiv
\frac{1}{2}\frac{\vartheta_3(\bar{q})^4}{\eta(\bar{q})^{12}}
\label{aepsilon}
\eea
The $\Gamma_{2,2}$ lattice has been written in (\ref{lattice}) as a
sum over all possible world-sheet instantons
\bea
\pmatrix{X^8\cr X^9}=
 W \pmatrix{\sigma^1\cr \sigma^2}\equiv
\pmatrix{m_1&n_1 \cr
m_2&n_2}\pmatrix{\sigma^1 \cr \sigma^2}
\label{M}
\eea
with worldsheet and target space coordinates $\sigma_1,\sigma_2$ and 
$X^8,X^9$ respectively, both taking values in the interval (0,1]. 
The entries $m_1,n_1$ are even or
odd integers depending on the specific orbifold sector, while $m_2,n_2$
run over all integers. We denote the three relevant sectors: $n_1$ odd,
$m_1$ odd and both odd by $\epsilon=+-,-+,--$ respectively.
The untwisted sector, $\epsilon=++$, will not contribute, since it has
too many zero modes to be soaked by the four vertex insertions at this order
in the momenta. Note that, due to the existence of these different sectors,
the duality group $\Gamma$ on $U$ is not the full $SL(2,Z)$ group but the
subgroup defined by the matrices 
\bea
\pmatrix{a & b \cr c & d}; ~~~~ b \in 2 {\bf Z} 
\label{Gamma}
\eea

The integration over the fundamental domain can be done
using the standard trick \cite{dkl}, which reduces the sum over all
wrapping mode configurations in a given $SL(2,{\bf Z})$ orbit to a
single integration over an unfolded domain. The unfolded domains are
either the strip or the upper half plane depending on whether
the orbit is {\it degenerate} ($ \det W=0$) or { \it non degenerate}
($\det W\neq 0$).
In \cite{bfkov,bgmn} these contributions were identified with the
perturbative and D-string instanton corrections respectively in the type I
side. The result for the non-degenerate
contributions were written as \cite{bgmn}
\bea
\langle (F_8)^\ell (F_9)^{4-\ell} \rangle_{nondeg}=
\frac{{\cal V}_{10}}{(4\pi^2\alpha^\prime)^4}t_8 F_8^\ell F_9^{4-\ell}
\frac{1}{T_2}
\sum_{n,\epsilon}\sum_{m_1,n_1,n_2}{\cal A}^n_\epsilon 
\frac{\partial^\ell}{\partial\nu_8^\ell}
\frac{\partial^{4-\ell}}{\partial\nu_9^{4-\ell}}I_n(\nu_8,\nu_9)
\label{at2}
\eea
where ${\cal A}^n_\epsilon$ are the coefficient in the $\bar{q}$
expansion of the modular forms (\ref{aepsilon}) and 
$I_n$ are the result of the modular integrations
\be
I_{n}=\frac{(U_2 T_2)^{\frac{1}{2}}}{m_1}
e^{-2\pi i T_1 m_1 n_2} 
e^{-2\pi i n(\frac{n_1+U_1 n_2}{m_1})}
e^{\pi i \nu_8(\frac{n U_2}{m_1 T_2}-m_1)}
\sqrt{\frac{\pi}{\beta}}e^{-2\sqrt{\beta\gamma}}
\label{Inf}
\ee 
with 
\bea
\beta&=&\pi(n_2^2 U_2 T_2+\nu_9 n_2 -\nu_8 U_1 n_2
-\frac{U_2\nu_8^2}{4 T_2})\nonumber\\
\gamma&=&\frac{\pi T_2}{U_2}(m_1+\frac{n U_2}{T_2 m_1})^2.
\eea
Here the sum over $m_1$ and $n_1$ are over even and odd integers, depending on
the $\epsilon$-sectors, and for a fixed $m_1$ the range of $n_1$ goes from $0$
to $m_1-1$.

In terms of the type I variables this result should come from 
D-instanton contributions since the duality relations
($T_2^F=T_2^I/\lambda_I$) implies that it goes like 
$e^{-\frac{N T_2^I}{\lambda_I}}$. Indeed it has been shown in 
\cite{bgmn} that its leading order in the $T_2$ expansion
\bea
&&\langle (F_8)^\ell (F_9)^{4-\ell} \rangle_{nondeg}=\nonumber\\
&&=\frac{{\cal V}_{10}}{(4\pi^2\alpha^\prime)^4}t_8 F_8^\ell F_9^{4-\ell} 
\frac{1}{T_2}
\sum_{\epsilon}\sum_{m_1,n_1,n_2}\frac{m_1^4}{m_1|n_2|}\frac{U^\ell}{U_2^4}
e^{-2\pi i T m_1 |n_2|}{\cal A}_{\epsilon}(\frac{n_1+\bar{U}n_2}{m_1})
\label{ffn}
\eea  
is exactly reproduced by a direct computation of D-instanton contributions 
in the type I side. In this identification the D-instanton
number $N=LM$ is mapped to the determinant $m_1 n_2$ of the representative matrices
$M$, while different wrapping mode configurations labelled by
$m_1,n_1,n_2$ are mapped to $L,s,M$,
describing the different sectors of 
the orbifold conformal field theory to which the effective 
$O(N)$ gauge theory flows in the infrared.
The coupling of the N D-instanton background to the $F_8,F_9$ gauge
fields is read from the classical D-instanton action
\be
S_{D-inst}=2\pi N T^F
-\frac{\pi N}{4\alpha^\prime U_2\lambda_I}
p_\nu(G_{\mu 9}+U G_{\mu 8})S_0^{cm}\gamma^{\mu\nu}S_0^{cm}+....
\ee
where $T^F=\frac{1}{\alpha^\prime}(B_{89}+\frac{i}{\lambda_I}\sqrt{G})$
and $S_0^{cm}= \frac{1}{N} \sum_{i=1}^N S^i$ are the fermionic
zero modes 
corresponding to the center of mass of the
N copies of D-string worldsheets. Indeed, bringing down four powers
of these terms, we soak the eight fermionic zero modes and combining
with the quantum D-instanton partition function 
$\frac{1}{N T_2}\sum_{M,L,s} M^{-4}{\cal A}(\frac{s+\bar{U}M}{L})$,      
the perturbative result (\ref{ffn}) is reproduced.  Note that the 
center of mass fermions here are defined with unconventional normalization in
order to make the comparison with the fundamental string side more transparent. In
this normalization
the partition function appears with $M^{-4}$ as stated above.
Had we used the canonical definition of
the center of mass fermion zero modes, then we would have had a factor $L^4$ 
instead of $M^{-4}$ in the
partition function. The easiest 
way to see the appearance of $L^4$ in the partition function is
by examining the short string sector $L=1, M=N$, where one can unambiguously
calculate the partition function using operator formalism. Since the $Z_N$
projection implies that the momenta $p_i$ of each string is equal (to $p$ say) 
the partition function appears with $e^{-NU_2 p^2}$ which upon integration over
$p$ yields a factor $1/N^4$. This cancels the factor $N^4$ coming from the trace
over canonically normalized fermionic zero modes (other than the center of mass
ones). In the general $(L,M)$ sectors the factor $L^4$ is then fixed by modular
transformations. Now one can rescale the canonically normalized zero modes by
$1/\sqrt{N}$ which gives rise to a jacobian $1/N^4$ giving rise to the partition
function claimed above.

The exact formula (\ref{at2}) shows the presence of $T_2$-subleading
terms which should correspond to a finite number of perturbative
corrections around the N D-instanton background. 
In order to make more transparent the translation in terms of the
type I variables it is convenient to use  the complex source
$\nu$, and its complex conjugate $\bar\nu$, defined as:
\be
\nu\equiv\frac{1}{2 U_2}(\nu_9+U \nu_8),
\ee
In terms of these new variables, after a long 
but straightforward algebra, one
can  express, to the order of interest, the $\bar\nu$-expansion of
the generating function ${\cal I}(\nu,\bar\nu)=\sum_n
{\cal A}_\epsilon^n I_n$ of (\ref{at2}) formally as:
\be
{\cal I}(\nu,\bar\nu)=
\sum_{r=0}^{4}\frac{\bar{\nu}^r}{r!} D^{r}_{\bar U} 
\left[ \frac{1}{(T_2-\frac{\nu}{2\pi})^{r+1}}
{\cal I}_0(\nu)  \right] +O({\bar\nu}^5),
\label{sublead}
\ee
where
\be
{\cal I}_0(\nu)=
\frac{{\cal V}_{10}}{(4\pi^2\alpha\prime)^4}
\sum_{\epsilon}\sum_{m_1,n_1,n_2}
\frac{e^{-2\pi i \bar{T} m_1 n_2}}{m_1 n_2^5}
{\cal A}_\epsilon(\frac{n_1+\bar{U} n_2}{m_1}) e^{2 m_1 n_2\nu}
\label{sublead0}
\ee
and $ D_{\bar U}$ is the $\bar U$-covariant derivative, which acting on a
modular form of weight $-2r$ gives a the weight $-2r+2$ form
\bea
D_{\bar U} \Phi_r=
(\frac{i}{\pi}\partial_{\bar{U}}-\frac{r}{\pi U_2}) \Phi_r
\label{cov}
\eea
In (\ref{sublead})  the first 4 terms
in the series have been explicitly indicated, 
being the only relevant ones to the present discussion,
The function appearing there has no
definite modular transformation properties, and the covariant
derivative ${D}^r_{\bar U}$
should be understood as acting on the coefficient of $\nu^{4-r}$ 
in the expansion of the function inside the brackets in (\ref{sublead}).
This gives the ${\bar F}^r F^{4-r}$ gauge couplings we are 
interested in, with $F=\frac{1}{2 U_2}(F_9+U F_8)$ and 
$\bar{F}=\frac{1}{2 U_2}(F_9+\bar{U} F_8)$:

\be
\langle {\bar F}^r F^{4-r} \rangle = 
\frac{{\cal V}_{10}}{(4\pi^2\alpha\prime)^4}\sum_{N} 
\frac{e^{-2\pi i N\bar{T}}}{N T_2^{r+1}}
\sum_{k=0}^{4-r} \frac{1}{(NT_2)^k}\frac{(r+k)!}{k!
(r!)^2 (4-r-k)!} H_N (D_{\bar{U}}^r A_{-+})
\label{sublead01}
\ee
where $H_N$ is the Hecke operator for the subgroup $\Gamma$ :
\begin{equation}
H_N (D_{\bar{U}}^r A_{-+}) = 
\sum_{\epsilon} \sum_{N|L}\sum_{s=0}^{L-1} L^{4-2r}   
D_{\bar{\cal{U}}}^r A_{\epsilon}(\bar{\cal{U}})|_{\bar{\cal{U}} =\frac{M\bar{U}
+s}{L}}
\label{hecke}
\ee
where $(L,s)$ are (even,odd), (odd,even) and (odd,odd) in the three $\epsilon$
sectors $+-$, $-+$ and $--$ respectively. Note that $D_{\bar{U}}^r A_{-+}(\bar{U})$
is a
modular form of weight $2r-4$ with respect to the subgroup $\Gamma$. The above
definition of the Hecke operator coincides with the standard one appearing for
example in \cite{gun} upto an overall $N$ dependent factor.

We would like now to show how these corrections 
can be reproduced by the effective D-string instanton action. 
To this aim, we start with the D-string instanton action in the 
presence of constant field strength background for the gauge
fields $G_{\mu i},B_{\mu i}$ under consideration. Each field strength appears in
the action with two fermion zero modes and therefore the amplitude is obtained
by extracting the 8 fermion zero modes in the partition function. One immediately
encounters two problems:

1) One needs to  keep in the action higher order terms in the fermion zero modes.
Unfortunately in curved backgrounds the covariant Green-Schwarz actions that exist
in the literature include only terms upto quartic in fermions, with higher order
terms in principle being calculable using space-time supersymmetry and kappa
symmetry. On dimensional grounds, as will become clear in the following, each pair
of fermion zero modes appears with $1/T_2$
and therefore, such higher order
terms can contribute to order $1/T_2^2$ correction (compared to the leading order
term). In the absence of information on such higher order terms we will be
restricted below to calculate
only the first subleading correction. However later we will argue that the complete
action should correctly reproduce the full perturbative expansion. 

2) Even upto the terms quartic in fermion fields, the Green-Schwarz action 
presupposes that the background fields satisfy the supergravity equations of motion.
Of course constant field strengths satisfy
the gauge field equations, however they can contribute to the stress energy
tensor and therefore can modify the classical equations for dilaton and the
8-dimensional 
metric. However this problem will not appear if we restrict ourselves to the first
subleading correction in
$1/T_2$. Indeed as seen from equation
(\ref{sublead01}), the first subleading term appears in two ways: 1) for
$\bar{\nu}=0$
(ie. $F_{\mu \nu z}=0$) there is a correction coming from the first order
expansion of
$1/(T_2-\frac{\nu}{2\pi})$. This can get contribution from terms in the action that
are quadratic in $F_{\mu \nu \bar{z}}$. For $F_{\mu \nu z}=0$ arbitrary
constant values of $F_{\mu \nu \bar{z}}$ will solve the equations of motion
since the resulting stress energy tensor is zero. 2) the first order in
$\bar{\nu}$ already comes with $1/T_2^2$ and therefore does not require an
expansion of the term $1/(T_2-\frac{\nu}{2\pi})^2$. This means that in the action 
we need to include only  the first order term in both the $F$'s which certainly
solves the 
linearized equations of motion. All the expessions in the following should be
understood to make sense only in the first subleading term in $1/T_2$.

The Green-Schwarz, covariant world-sheet action for a single 
string coupled to the ${\cal N}=1$, 10-D supergravity multiplet, is given by 
\cite{ft} ($\alpha^\prime=\frac{1}{2}$):  
\bea
S_{D-inst}=2\pi\int d^2 \sigma
\left[\frac{1}{\lambda_I}\sqrt{|det \hat{G}_{a b}|}+\frac{1}{2}
\epsilon^{ab} R_{ab}\right] + \dots
\label{d-action}
\eea
The dots above indicate terms involving 6 and higher powers of fermion fields. Here
we have directly rewritten the action of \cite{ft}
in the Nambu-Goto form, by
using the equations of motion for the world-sheet metric. The resulting
induced metric ${\hat G}_{ab}$ and the antisymmetric coupling $R_{ab}$ are
given by:
\bea
\hat{G}_{a b}&=&\partial_{(a} X^M \partial_{b)} X^N 
G_{M N}-2i\partial_{(a} X^M \bar{\Theta}\gamma_M D_{b)} \Theta\nonumber\\
&&-\bar{\Theta}\gamma_M D_a \Theta\bar{\Theta}\gamma^M D_b \Theta    
\nonumber\\
R_{a b}&=&i\partial_{[a} X^M \partial_{b]} X^N 
B_{M N} + \frac{2i}{\lambda_I}\partial_{[a} X^M 
\bar{\Theta}\gamma_M\sigma_3 D_{b]}
\Theta\nonumber\\
&&+\frac{1}{\lambda_I}\bar{\Theta}\gamma_M D_{[a}
\Theta\bar{\Theta}\gamma^M\sigma_3 D_{b]} \Theta. 
\label{gab}
\eea
Here the covariant derivative $D_a$ is defined to be:
\be
D_a=\partial_a-\frac{1}{4}\gamma^{\hat{K}\hat{L}}
\tilde{\omega}_N^{\hat{K}\hat{L}}\partial_a X^N,
\label{cd}
\ee
where
$\tilde{\omega}_N^{\hat{K}\hat{L}}=
\omega_N^{\hat{K}\hat{L}}+\frac{\lambda_I}{2}H_N^{\hat{K}\hat{L}}\sigma_3$,
with $\omega$ 
the spin connection
and $H$ the 3-form field strength of the RR antisymmetric tensor $B$.
The capital, latin indices are 10-dimensional 
curved indices, whereas hatted ones are $SO(1,9)$ flat
indices. $\Theta^T=(\theta,\tilde{\theta})$ denotes a doublet of $SO(1,9)$
Majorana-Weyl spinors on which $\sigma_3$ acts.
$\theta$ and $\tilde{\theta}$ have the same $SO(1,9)$ chirality and
are world-sheet scalars.
Finally, $\bar\Theta\equiv (\theta^T\gamma^0,\tilde{\theta}^T\gamma^0)$.
Note that the fields appearing above are supposed to satisfy the 10-dimensional
supergravity equations of motion.

We then proceed by choosing the static gauge, which amounts
to identifing the euclidean world-sheet complex coordinates  
$z\equiv \frac{1}{\sqrt{2\tau_2}}(\sigma^1+\tau\sigma^2)$, 
$\bar{z}$,  with the spacetime torus coordinates 
$X^z\equiv\frac{1}{\sqrt{2\tau_2}}(X^8+U X^9)$, $X^{\bar{z}}$
respectively and the complex structure of the world-sheet 
torus, $\tau$, with that of the spacetime torus, $U$.  
In addition, local kappa-symmetry 
allows to reduce the fermionic degrees of freedom
by imposing \footnote{Strictly speaking, because of the Majorana-Weyl nature of
fermions, the gauge fixing condition makes sense only in the Minkowski 
space-time
(and world-sheet). We assume that $X^9$ is time like in which case 
$z$ and $\bar{z}$
are real light-like coordinates and $\tau_2$ and $U_2$ are imaginary. 
At the very end we will
do the analytic continuation back to complex $\tau$ and $U$}
\bea
\gamma^z\theta &\equiv& \frac{1}{\sqrt{2U_2}}(\gamma^8+U\gamma^9)\theta
=0\nonumber\\
\gamma^{\bar{z}}\tilde{\theta} &\equiv& 
\frac{1}{\sqrt{2U_2}}(\gamma^8+\bar{U}\gamma^9)\theta
=0
\eea 
The resulting gauge fixed fermions
$\theta$ $(\tilde{\theta})$ become left-moving (right-moving)
world-sheet fermions in the $\bf 8_s$ $(\bf 8_c)$ of $SO(8)$.

We will be interested in a constant background for the field strength
of $A_{\mu i}\equiv \frac{G_{\mu i}}{\lambda_I}+ B_{\mu i}$\footnote{Notice
that precisely this combination
appears in the connections in (\ref{cd})}. To extract
the relevant couplings it is convenient to
choose  for the vielbeins the representation
\bea
e_M^{\hat{L}}=\pmatrix{e_i^j&G_{\mu i}\cr 0& \eta_{\mu\nu}}+
O(G_{\mu i}^2)
\label{e}
\eea
with $e_i^j$ the square root of the torus metric (in the complex
basis $e_z^{z}=e_{\bar{z}}^{\bar{z}}=\sqrt{T_2^I}$,
$e_{\bar{z}}^{z}=e_z^{\bar{z}}=0$). 
Since we are interested
in extracting four powers of the field strengths $A_{\mu i}$ from the expansion of 
(\ref{d-action}) and then set $G_{\mu i}$ to zero, we can set
always $G_{\mu i}$ to zero unless a derivative hits on it. This is so because
such terms would anyway disappear in the final result due to gauge invariance.
Furthermore since the connections involve at most one derivative of the metric,
the terms containing
two or more derivatives of $G_{\mu I}$ are irrelevant for our present
discussion. Here and in the 
following we will always set these components to zero. 

Using the ansatz (\ref{e}), and recalling from \cite{bgmn}
that, after orientifolding, $\theta$ is periodic (and therefore 
has 8 zero modes, denoted by  $\theta_0$), whereas $\tilde{\theta}$ is
anti-periodic, it can be seen that the contributions of interest  
in (\ref{gab}) come from two kind of terms,
$\bar{\theta_0}\gamma^{\bar{z}}\gamma^{\hat{K}\hat{L}}
\tilde{\omega}_{i}^{\hat{K}\hat{L}}\theta_0$ for $i=z,\bar{z}$ and
$\bar{\theta}_0\gamma^\mu\gamma^{\hat{K}\hat{L}}
\tilde{\omega}_{\rho}^{\hat{K}\hat{L}}\theta_0$.
The first type of terms gives the following fermionic bilinears 
\bea
\eta&=&\frac{i}{4}\partial_{[\mu}
A_{\nu]\bar{z}}\bar{\theta}_0\gamma^{\mu\nu \bar{z}}\theta_0\nonumber\\
\bar{\eta}&=&\frac{i}{4}\partial_{[\mu}
A_{\nu] z}\bar{\theta}_0\gamma^{\mu\nu \bar{z}}\theta_0,
\label{eta}
\eea
whereas the second type gives
\be
\epsilon_{\rho\mu} = \frac{i}{4}\partial_{[\nu}
A_{\rho] \bar{z}}\bar{\theta}_0\gamma^{\mu\nu \bar{z}}\theta_0.
\label{epsi}
\ee
In the above formulae and in the following  $\gamma_{z}$ and
$\gamma_{\bar{z}}$ denote $SO(2)$ gamma matrices and the $T_2$ dependence will be
explicitely displayed. 
Contributions to the ${\cal F}^4$ couplings will come from
the  D-instanton partition function once four powers of   
these combinations are brougth down to soak the 8 fermionic zero  
modes. This is why we replaced the fermions in (\ref{eta},\ref{epsi}) by their
zero mode part $\theta_0$. ${\cal F}^4$ couplings will be defined
then by the fourth order terms in the $\eta$, $\bar\eta$ expansion 
of the resulting effective action. 
The two pieces in the lagrangian(\ref{d-action}) can then be written as
\footnote{We omit in this expression a trivial 
$\lambda_I$-rescaling of all bosonic $X_\mu$ and 
fermionic fields $\theta,\tilde{\theta}$}
\bea
\frac{1}{\lambda_I}\sqrt{det\, \hat{G}_{ab}}&=&T^F_2-\eta
+\partial_z {X}^\mu\partial_{\bar{z}}{X}_\mu
-i\bar{\tilde{\bf\theta}}\gamma_{\bar{z}}\partial_{z}
\tilde{{\bf \theta}}
-i\bar{\theta}\gamma_z\partial_{\bar{z}}\theta+...\nonumber\\
&&+\frac{\bar{\eta}}{T^F_2-\eta} (\partial_{\bar{z}} {X}^\mu
\partial_{\bar{z}}{X}_\mu- i
\bar{\tilde{{\bf \theta}}}\gamma_{\bar{z}}
\partial_{\bar{z}}\tilde{{\bf \theta}})
+...\nonumber\\
\frac{1}{2}\epsilon^{ab} R_{ab}&=&-i B_{z\bar{z}}-\eta
-i\bar{{\bf \tilde{\theta}}}\gamma_{\bar{z}}\partial_{z}{\bf \tilde{\theta}}
-i\bar{\theta}\gamma_z\partial_{\bar{z}}\theta\nonumber
-i\frac{\bar{\eta}}{T^F_2-\eta}\bar{\tilde{\theta}}\gamma_{\bar{z}}
\partial_{\bar{z}}{\tilde{\theta}}+...
\label{gr}
\eea
where $\dots$ represent higher quantum fluctuations,
and we have rescaled  ${X}^\mu$, $\tilde{{\bf\theta}}$ as
\bea
&&{X}^\mu \rightarrow
(1-\frac{2\epsilon}{T^F_2})_{\mu\nu}X^{\nu}~~~~\mu=0,\dots 7\nonumber\\
&&\tilde{{\bf\theta}}_\alpha \rightarrow
(1-\frac{\eta}{T^F_2})^{\frac{1}{2}}{\tilde{\theta}}_\alpha
~~~~\alpha=1,\dots 8.
\label{scaling}
\eea  

We have considered so far the Nambu-Goto
action for a single D-string. The results (\ref{gr}) can however
easily be generalized to the N D-string case.  
The low energy effective action describing the excitations of the N 
D-string system is described by $O(N)$ gauge theory studied in
\cite{bgmn}. As argued in that reference, after integrating out the 
very massive degrees of freedom in the infrared limit, we are
left with an effective conformal field theory in terms of the
diagonal multiplets $X_\mu^i$, $\theta^i$ and $\tilde{\theta}^i$,
$i=1,\dots,N$, on which the orbifold permutation group
$S_N$ is acting. Corespondingly, in (\ref{gr})
the fermionic bilinears $\eta$ and $\bar\eta$ will
involve 
$\sum_{i=1}^N\bar\theta^i\gamma^{\mu\nu\bar{z}}\theta^i=
N\bar{\theta}_0^{cm}\gamma^{\mu\nu\bar{z}}\theta_0^{cm}+...$, $\theta_0^{cm}$
being the fermionic zero mode corresponding to the center of mass.
Also, the volume factor 
$\sum_{i=1}^N G_{z\bar{z}}\partial_z  X_i^z \partial_{\bar{z}}X_i^{\bar{z}}=
N T_2^I$
pick up a factor of N. The effective action is then given by simply
replacing $T^I_2, \eta,\bar{\eta}$ and $\epsilon$  by $N T^I_2, 
N\eta,N\bar{\eta}$ and $N\epsilon$. 
The
operators expressed in terms of the  non-zero mode fields will
be replaced by the sum of the $N$ copies of them, involving $X_\mu^i$
$\theta^i$ and $\tilde{\theta}^i$.   
The effective action then reads
\bea
S_{ND-inst}&=&2\pi i N T^F-4\pi N\eta+2 \pi\sum_{i=1}^N \int d^2 z  
\left[ \partial_z {X}^{\mu i}\partial_{\bar{z}}{X}^i_{\mu}
-2i\bar{\tilde{{\bf \theta}}}^i\gamma_{\bar{z}}\partial_{z}
\tilde{{\bf\theta}}^i\right.\nonumber\\
&&-2i\bar{\theta}^i\gamma_z\partial_{\bar{z}}\theta^i
+\left.\frac{\bar{\eta}}{T_2-\eta} (\partial_{\bar{z}}{X}^{\mu i}
\partial_{\bar{z}}{X}^i_\mu-2 i
\bar{\tilde{\bf\theta}}^i\gamma_{\bar{z}}
\partial_{\bar{z}}\tilde{{\bf\theta}}^i)
+...\right]
\label{actionND}
\eea
The moduli dependence of ${\cal F}^4$ couplings in the N D-instanton 
background are then defined by the $\eta^{4-r}\bar{\eta}^r$ terms in the
expansion of the exponential of (\ref{actionND}).
Identifying the fermionic bilinears $2\pi\eta,2\pi\bar{\eta}$ with our previous
sources  $\nu,\bar{\nu}$, one can show that the         
perturbative results (\ref{sublead}), (\ref{sublead0}) are reproduced.

Let us start by considering the $\bar{\eta}=0$ case.
Were it not for the rescaling of the $X^{\mu},\tilde{\theta}$ fields
(\ref{scaling}),
we would have a free theory (upto Weyl permutations)
defining the orbifold partition function 
$\frac{1}{N T_2}\sum_{M,L,s} M^{-4}{\cal A}(\frac{s+\bar{U}M}{L})e^{2N\eta}$ 
\cite{bgmn,gmnt}. 
Identifying as before $L,M,s$ with $m_1,n_2,n_1$ respectively we can see
by a simple inspection of (\ref{sublead}), (\ref{sublead0})
that we reproduce the $\bar{\eta}=0$ term      
except for the fact that we get $\frac{1}{T^F_2}$ instead of 
$\frac{1}{T^F_2-\eta}$ that appears in (\ref{sublead}). 

We want now to show that,
at least to first order in $\frac{1}{T_2^F}$, 
the rescaling of the fields
(\ref{scaling}) results in the desired renormalization of the inverse
area prefactor $\frac{1}{T^F_2}\rightarrow   
\frac{1}{T^F_2-\eta}$.   
Indeed, to the first order, the Jacobian of the transformation (\ref{scaling})
is given by the following correlation function: 
\be
\frac{2\pi}{T_2^F}\int d^2z~\left[ 4 \epsilon_{\mu\nu} \langle\partial_z X^\mu
\partial_{\bar{z}} X^\nu\rangle-2i \eta \langle\bar{\tilde{\theta}}
\gamma^z\partial_z\tilde{\theta}\rangle\right].
\label{corr}
\ee 
This expression, which involves expectation values
of the kinetic energies of $X^\mu$ and $\tilde{\theta}$,
needs to be regularized. If we adopt a point-splitting 
regularization, the term involving $\tilde{\theta}$ is zero
because $\tilde{\theta}$ has no zero modes. We are thus left
with the bosonic contribution, which can be evaluated using:
\be
\langle\partial_z X^\mu (z)\partial_{\bar{z}} X^\nu (w)\rangle=
-\frac{\delta^{\mu\nu}}{8\pi}.
\ee
The claimed result follows after performing the $z$-integration
and using $\delta^{\mu\nu}\epsilon_{\mu\nu}=-\eta$.
We expect that higher order terms reproduce 
the expansion of $\frac{1}{T^F_2-\eta}$.
We can then write finally 
\bea
{\cal I}^{Dinst}_0(\eta) &\equiv& \langle e^{-S_{ND-inst}(\eta,0)}\rangle
\nonumber\\
&=&\frac{{\cal V}_{10}}{(2\pi^2)^4}\sum_{\epsilon}\sum_{L,M,s}
\frac{1}{LM^5 (T^F_2-\eta)}
e^{-2\pi i \bar{T_F} LM}{\cal A}_\epsilon(\frac{s+\bar{U} M}{L})
e^{4\pi LM \eta}
\label{Id}
\eea
that reproduces the $r=0$ term in (\ref{sublead}) after the previous
identifications. 

We can now go on and consider $r$ $\bar{\eta}$ insertions in (\ref{Id}). From 
the $\bar{\eta}$ coupling in (\ref{actionND}) 
one can see that each of these correspond to the insertion of a normalized
stress-energy tensor $\frac{1}{T^F_2-\eta}\int d^2 z \overline{T}(\bar{z})$ with
${\overline T}(\bar z)\equiv
\sum_{i=1}^N \left[
\partial_{\bar{z}}X^{\mu i}\partial_{\bar{z}}X^i_\mu(\bar z)-
2i\bar{\tilde{\theta}}^i
\gamma_{\bar{z}}\partial_{\bar{z}}\tilde{\theta}^i(\bar z)\right]$.
But this is precisely the Virasoro generator of an infinitesimal shift 
in the $\bar{z}$ worldsheet coordinate, ${\bar L}_0$, which couples
to $\bar U$. Therefore its insertions translate
into covariant derivatives $D_{\bar{U}}$ acting on (\ref{Id}). Note that although
the expressions above were to make sense only to the first subleading term in
$1/T_2$ the fact that the $\bar{\eta}$ coupling appears with the same normalization
factor $\frac{1}{T^F_2-\eta}$ as in the perturbative result (\ref{sublead}) is
quite remarkable.

The calculation above shows the precise matching of the first subleading correction
around D-string instantons with the exact result on the fundamental string side. How
can we extend this to all orders in $1/T_2^F$ ? The steps involved are two fold.
First
of all, we will need terms upto 8th order in fermion zero modes in the covariant
Green-Schwarz action. This can in principle be obtained by repeatedly using 
space-time supersymmetry and kappa symmetry. Secondly, we need to solve the
linearized equation for $\bar{\eta}$ in the presence of arbitrary $\eta$. This will
involve turning on 8-dimensional gravitational and dilaton fields to the first order
in $\bar{\eta}$. Although straightforward, the computation involved is rather
cumbersome. We will give here an alternative argument to show that the complete
action should correctly reproduce the full perturbative expansion. For $N=1$ the
covariant D-string action is the same as the fundamental string action. The static
gauge
computation presented here should be the same as the light cone gauge computation
in the fundamental string side provided one restricts to $det W=1$ in equation
(\ref{sublead01}). This shows that at least for single instanton the two results should
agree to all orders in $1/T_2$. For $N > 1$, the orbifold CFT description implies
that the result is obtained from the $N=1$ result by applying the Hecke operator 
$H_N$, upto $N$ dependent factors. Indeed, by the zero mode argument of
\cite{gmnt,bgmn}, one knows that only the twisted sectors of the form $(L)^M$
contribute and the resulting expression involves the sum over $L,M,s$. Modular
covariance in $\bar{U}$ under the subgroup $\Gamma$ then fixes the powers of $L$
in the sum over the sectors, giving rise to Hecke operator $H_N$. The only quantity
which is not fixed in the amplitude $\langle {\bar F}^r F^{4-r} \rangle$ by this 
argument is
the $N$ dependence of the leading as
well as the
subleading terms in $1/T_2^F$ for each $r$. To determine this $N$ dependence we note
that the general form of the $N$ D-string action in the orbifold limit is 
schematically of the form
\bea
S_{ND-inst}&=&  NT_2^F\sum_{k=0}^{4} \frac {a_k}{(T_2^F)^k} (F_{\bar{z}}
\theta_0^2)^k  
+\sum_{k=0}^4 \frac {1}{(T_2^F)^k} (F_{\bar{z}} \theta_0^2)^k \sum_{i=1}^N A_k(X^i,
\theta^i,
\tilde{\theta}^i)\nonumber\\
&&+ \frac{1}{T_2^F}(F_z \theta_0^2) \sum_{k=0}^3 \frac {1}{(T_2^F)^k} (F_{\bar{z}}
\theta_0^2)^k 
\sum_{i=1}^N B_k(X^i, \theta^i,
\tilde{\theta}^i)
\label{sg}
\eea
where $a_k$ are numerical coefficients independent of $N$ and $A_k$ and $B_k$
are dimension (1,1) and (0,2) operators (depending on $X$, $\tilde{\theta}$ and
the non-zero modes of $\theta$) for each of the $N$ copies of the variables. Here
we have suppressed all 8-dimensional gamma matrices and Lorentz indices. 
Note that (\ref{actionND}) obtained by including only upto quartic terms in fermion
fields in the
covariant Green-Schwarz action fixes $a_0, a_1, a_2, A_0, A_1$ and $B_0$ and these
were the objects that entered in the computation of the first subleading
corrections. In writing (\ref{sg}) 
we have assumed that the higher dimensional operators will decouple in the
infrared limit. The powers of $T_2^F$ in the above are easily seen by scaling
arguments
while the $N$ dependence follows from the definition of the fermion zero mode
$\theta^{cm} =\frac{1}{N} \sum_i \theta^i$.

One can see from this general form of the action, that the leading term
for each $r$
comes with $\frac{1}{N} H_N (D_{\bar{U}}^r A_{-+})$ while in the subleading
correction in
$1/T_2^F$, the volume $T_2^F$ is replaced by $NT_2^F$. This exactly reproduces the
exact
formula (\ref{sublead01}) of the fundamental string. 

To summarize, we have studied the perturbative corrections around D-string
instantons and have shown that there is an expansion similar to the one appearing on
the fundamental string side. In particular by an explicit analysis of the
supersymmetric Nambu Goto action in static gauge that appears in the orbifold CFT,
the first subleading term was shown to agree precisely on the two sides. We further
argued by studying the $N$ dependence of the general form of Nambu Goto action
containing higher powers of fermion zero modes, that the complete perturbative
expansion is reproduced by the D-string instantons. Thus our results show that the
orbifold CFT description of the D-string instantons captures also these perturbative
corrections. 

The results presented here are restricted to the simplest example of eight
dimensional dual pairs with sixteen supercharges, among the ones studied in
\cite{bfkov}, \cite{bgmn}. This model is however a convenient prototype for all of
them (type I-heterotic, CHL-type I with non commuting Wilson lines, etc.)  and we
believe that these results extend in a straightforward way to all the cases in which
the precise matching between fundamental and D-string BPS excitations has been
stablished. 

The fact that we can compute perturbative corrections around D-string instanton
backgrounds by inserting operators that are not of dimension (1,1) (in the present
context the stress-energy tensor $\overline{T}(\bar{z})$) is quite remarkable and we
believe it deserves a deeper study. 

As in the previous studies \cite{bfkov}, \cite{bgmn}, this analysis was restricted
to $D>4$.  For $D\leq 4$ the five-brane physics becomes relevant. This is of course
another exciting direction to explore.

\rnc{\Large}{\normalsize}

\end{document}